\documentclass[]{raa}           

\usepackage{graphicx,times}
\usepackage{natbib}
\usepackage{amssymb,amsmath}
\bibpunct{(}{)}{;}{a}{}{,}

\usepackage[pagebackref=true]{hyperref}

\begin{document}

\title{The Tianlin Mission: a 6m UV/Opt/IR space telescope to explore the habitable worlds and the universe}

\volnopage{ {\bf 20XX} Vol.\ {\bf X} No. {\bf XX}, 000--000}
\setcounter{page}{1}

   \author{Wei Wang\inst{1}, Meng Zhai\inst{2}, Gang Zhao\inst{1,3}, Shen Wang\inst{1}, Jifeng Liu\inst{1}, Jin Chang\inst{1}, Xuejun Zhang\inst{4}, Jihong Dong\inst{4}, Boqian Xu\inst{4}, Frank Grupp\inst{5,6}
   }
%\noalign{\smallskip}

   \institute{CAS Key Laboratory of Optical Astronomy, National Astronomical Observatories, Chinese Academy of Sciences, Beĳing 100101, PR China; {\it wangw@nao.cas.cn, mzhai@nao.cas.cn}\\
        \and CAS South America Center for Astronomy, National Astronomical Observatories, Chinese Academy of Sciences, Datun Road A20, Beijing 100101, China\\
        \and
             University of Chinese Academy of Sciences, Beijing 100049, China\\
	\and
	  The Changchun Institute of Optics, Fine Mechanics and Physics, Chinese Academy of Sciences\\
	  \and
Universitäts Sternwarte München, Scheinerstr 1, München 81679, Germany\\
\and 
Max Planck Institute for Extraterrestrial Physics, Giessenbachstrasse 1, 85748 Garching, Germany\\
\vs \no
   {\small Received 20XX Month Day; accepted 20XX Month Day}
}

\abstract{It is expected that the ongoing and future space-borne planet survey missions including TESS, PLATO, and Earth 2.0 will detect thousands of small to medium-sized planets via the transit technique, including over a hundred habitable terrestrial rocky planets. To conduct a detailed study of these terrestrial planets, particularly the cool ones with wide orbits, the exoplanet community has proposed various follow-up missions. The currently proposed ESA mission ARIEL is the first step for this purpose, and it is capable of characterization of planets down to warm super-Earths mainly using transmission spectroscopy. The NASA 6m UV/Opt/NIR mission proposed in the Astro2020 Decadal Survey white paper is to further tackle down to habitable rocky planets, and is expected to launch around 2045. In the meanwhile, China is funding a concept study of a 6-m class space telescope named Tianlin (A UV/Opt/NIR Large Aperture Space Telescope) that aims to start its operation within the next 10-15 years and last for 5+ years. Tianlin will be primarily aimed to the discovery and characterization of rocky planets in the habitable zones (HZ) around nearby stars and to search for potential biosignatures mainly using the direct imaging method. Transmission and emission spectroscopy at moderate to high resolution will be carried out as well on a population of exoplanets to strengthen the understanding of the formation and evolution of exoplanets. It will also carry out in-depth studies of the cosmic web and early galaxies, and constrain the nature of the dark matter and dark energy. We describe briefly the primary scientific motivations and main technical considerations based on our preliminary simulation results. We find that a monolithic off-axis space telescope with primary mirror diameter larger than 6m equipped with a high contrast chronograph can identify water in the atmosphere of a habitable-zone Earth-like planet around a Sun-like star. More simulations for the detectability of other key biosignatures including O3, O2, CH4, and Chlorophyll are coming.
}

   \authorrunning{W. Wang et al. }            %author_head in even pages
\titlerunning{The Tianlin Mission and Water Detection}  % title_head in odd pages
%\linenumbers
%\doublespacing
\maketitle

%________________________________________________ sections below
% 
\section{Introduction}
\label{sect:intro}  % \label{} allows reference to this section

The first extrasolar planet (exoplanet) orbiting a solar-type star was discovered in 1995~\citep{MQ1995} via the radial velocity (RV) method, which is to measure precisely the periodic RV variation of a star, with the RV amplitude varying from about 100~m/s for Jupiter-like planets to a few m/s for Earth-like planets around M type stars. Since then, 5084 exoplanets have been detected and confirmed\footnote{https://exoplanetarchive.ipac.caltech.edu/index.html accessed on September 5, 2022}, among which most are discovered via the RV and transit methods. While the RV technique is the most efficient one before 2009, the transit method, which is to detect planets by searching for small dimming in the light curve when a planet transits and blocks a small portion of its host star's disk, has turned out to be much more efficient in revealing planet candidates. The {\textit Kepler} mission \citep{Borucki2011a} and the K2 mission \citep{Howell2014} have opened a new era with thousands of exoplanet detected in less than 10 years, bringing up a lot in new knowledge of exoplanets, particularly of small and rocky planets. 

The Transiting Exoplanet Survey Satellite (TESS) project~\citep{Ricker2015} has been monitoring approximately 200, 000 stars brighter than 12 mag in $V$, and it is expected to discover more than 14,350 planets including over 2100 planets smaller than 4$R_\oplus$ and 280 smaller than 2$R_\oplus$. About 70 habitable zone (HZ) planets will be detected but should be mostly orbiting M dwarfs~\citep{Barclay2018}. As of September 2020, TESS finishes its primary mission, with 2174 candidate planets and 67 confirmed planets reported. It is recently approved to continue working until 2022 and should find more planets than the original predictions. The PLATO space mission~\citep[PLAnetary Transits and Oscillation of stars; ][]{Rauer2014} is to be launched in 2026, which will take a further step to search \textcolor{blue}{for} small and cool planets. It will stare on two 2232~deg$^2$ sky areas for 2-3 years, aiming to detect terrestrial exoplanets at orbits up to the HZs of solar-type stars with magnitude between 4-16 mag \citep{Rauer2014}. It will observe up to 1 million stars, and may detect 13,000 planets, including 2,000 planets smaller than 2$R_\oplus$, and 6-280 of them in HZs. A new survey mission named Earth 2.0 (or ET in short) has been recently proposed in China, which aims to detect a dozen Earth-like planets around solar-type stars, and thousands of other types of planets from sub-Earths to Jupiters. ET is planned to launch in 2026 and operate for 4 years or more~\citep{Ge2022}. In addition to the above-mentioned transit survey missions, a space-borne astrometric mission named Closeby Habitable Exoplanet Survey (CHES) is proposed for the detection of habitable planets of the nearby solar-type stars~\citep{Ji2022}.

Therefore, various missions in the 2020s may discover thousands of rocky planets orbiting stars that are bright enough for ground-based RV follow-up confirmations, including tens or hundreds of temperate terrestrial planets with liquid water pooling on their surfaces. Only very few of the systems, with nearby cool host stars, relatively large planet radius, and short orbiting period, have been preliminarily investigated by currently available telescopes \citep{Benneke2019,Swain2021}, while most of them have to wait for future missions for atmospheric characterization. 

The ARIEL (Atmospheric Remote-sensing Infrared Exoplanet Large-survey) mission \citep{Tinetti2016} is selected by the European Space Agency (ESA) as the fourth medium-class mission planned for launch in 2028. The goal of the ARIEL mission is to obtain spectra of $\sim$1000 known planets orbiting nearby stars using the transit method, to study and characterize the planets' chemical composition and thermal structures, in order to address the fundamental questions of how planetary systems form and evolve. ARIEL will mainly focus on gaseous and rocky planets with equilibrium temperatures higher than 500\,K. Given its meter-class aperture, it is not capable of investigating the atmospheres of habitable terrestrial planets around GK type stars, unfortunately. 

One of the key scientific objectives of the James Webb Space Telescope ~\citep[JWST;][]{Gardner2006} is to determine the physical and chemical properties of planetary systems, and investigate the potential for the origins of life in those systems. However, there is no spectrograph equipped behind its coronagraph, which means it does not provide the capability of obtaining planet spectra directly, and transit spectroscopy at low-to-intermediate resolution (i.e., R=$150-4000$) is the only technique to obtain spectra of exoplanets by JWST. In this case, terrestrial planets in the HZs of FGK stars more than 3 pcs away could barely be investigated in less than 100 hours, simply because the star-to-planet photon ratio is overwhelming large which makes it almost hopeless to build enough signal-to-noise ratio (SNR) for planet spectra in its mission lifetime. However, there is a lack of HZ terrestrial planets around nearby FGK stars that can not be made up for in the near future.

The 2.4m Nancy Grace Roman Space Telescope (or Roman in short) is NASA's next flagship observatory, planned to launch no later than 2027. The Coronagraph Instrument (CGI) on RST will be the first space coronagraph that aims to demonstrate the high-contrast technology necessary for exoplanet imaging and spectroscopy. The predicted contract ratio of CGI is below 10$^{-8}$ and close to 10$^{-9}$, and thus is about 2 to 3 orders of magnitude better than any current facility, and should allow direct imaging and spectroscopy of hot to cold Jupiters~\citep{Kasdin2020}.

The CGI's on-board experience within this decade should shed crucial light to future larger missions, including Habitable Exoplanet Observatory~\citep[HabEx;][]{Gaudi2020}, Large UV/Optical/Infrared Surveyor~\citep[LUVOIR;][]{LUVOIR2019}. An intermediate mission of these two is prioritized in the Astro2020 Decadal Survey and is to be maturated in the coming years and to be launched in the first half of the 2040s. It would provide opportunities to discover and characterize habitable Earth-like planets since then if the mission develops smoothly.

The next generation ground-based extremely large telescopes (ELTs) including the Thirty Meter Telescope (TMT), the Extremely Large Telescope (ELT) and the Giant Magellan Telescope (GMT) are expected to have their first light in 2027 or later. High-contrast coronagraphic instruments aided with extreme adaptive optics (ExAO) systems may be built and operating by 2035, and they may help to characterize spectrally warm giant planets with a contrast ratio of $\sim 10^{-8}$. It is worth noting that the high-dispersion spectroscopy \citep{snellen2014}, when combined with high-contrast imaging, i.e., HCI+HDS, using advanced coronagraph on the ELTs may allow atmospheric studies of a small sample of habitable-zone Earth-like planets around nearby red dwarfs~\citep{Snellen2015, WangJ2017}.

To conclude, there is currently no approved or planned mission before the 2040s that will be able to characterize the atmospheres of the temperate rocky planets around GK stars. Foreseeing the lack of the key capability to detect and identify extraterrestrial bio-signatures within the next decade, the Chinese Academy of Sciences (CAS) is funding a concept study for a 6m space telescope named Tianlin since 2019, aiming to characterize habitable planets around nearby GK type stars~\citep{Wang2020SPIE}. 

This paper is organized as follows. We present in Section~\ref{sect:obj} an overall introduction of Tianlin and its main science objectives, and summarize the currently known potential biosignature and our target molecular species in Section~\ref{sect:bios}. Section~\ref{sect:model} introduces the model and methods that we adopt for simulation. Constraints to technical requirements based on our simulation are presented and discussed in Section~\ref{sect:result}, following by a brief summary in Section~\ref{sect:summary}.

\section{Tianlin and its Scientific Objectives}
\label{sect:obj}

The Tianlin mission is a proposed space telescope with an aperture size of 4-6 meters launched to the Sun-Earth L2 point(SEL2) halo orbit. Its primary objective is the characterization of rocky planets, particularly HZ terrestrial planets around nearby G and K type stars. The telescope is considered to be an off-axis 3-mirror system with a primary mirror of 4-6 meters, including four Fine Guiding Sensors to aid high accuracy and high stability line-of-sight pointing, an advanced coronagraph, one backend spectrograph with resolution varying from $\sim$100 to 15,000. The main techniques to be used include direct spectroscopy via the coronagraph and transit spectroscopy with low to high resolution spectrograph, covering a full range from the UV to near-infrared (0.20-1.7$\mu$m). Like the similar missions including HabEx and LUVIOUR, the SEL2 quasi-halo orbit is considered for Tianlin, for the consideration of the stable dynamical, thermal environment and the capability of continuous staring for tens to hundreds of hours. This orbit can be reached by the Chinese Long March 5 rocket, and was visited once by the Chinese Chang'e-2 mission in 2012.

Tianlin is different from HabEx in several aspects. First, the aperture size of Tianlin may be as large as 6m. Secondly, Tianlin will equip a high-resolution spectrograph to utilize the high-resolution spectroscopic atmospheric studies and enable the HCI+HDI technique to better trace spectral signals, while HabEx will not. Lastly, a formation-fly starshade configuration is not considered for Tianlin at the current stage. The NASA IROUV mission is a hybrid of HabEx and LUVOIR named Habitable Worlds Observatory (HWO), and it aims to directly image planets including Earth-like exoplanets around Sun-like stars and characterize their atmospheres. This mission is to be matured in the next 5 years, and may be launched in the first half of the 2040s.

The mission has a primary science objective to search and characterize the atmospheres of nearby exoplanets, particularly Earth-like planets and close-in rocky planets around G and K stars, to explore their habitability, and to search for potential biosignatures in their atmospheres or on their surfaces. We define in this work an Earth-like planet as a rocky planet with a radius between 0.6 to 1.4$R_\oplus$ orbiting around a G or K-type star within the respective conservative HZs following the work by \cite{Kopparapu2014}. We define a twin Earth as a rocky planet with a radius of 1$R_\oplus$ orbiting a G2 star at 1~AU orbit. A close-in rocky planet in our sample is a planet that orbits a G or K dwarf star between the inner limits of the conservative and optimistic HZs. It may also be habitable, and should be easier for atmospheric characterization compared to Earth-like and Earth-twin planets. 

The detailed target sample for this objective is to be built up in the following years, based on the transit and direct imaging surveys of nearby stars. It will consist not only of transiting planets but also of non-transiting planets. As summarized in Section~\ref{sect:intro}, there will be plenty of transiting Earth-like planets to be found by the above-mentioned transit surveys, for which Tianlin could only obtain their transmission or reflection/emission spectra during their primary/secondary transits without blocking central stars' light. However, as will be emphasized in Section~\ref{sect:result}, such a technique may need the aperture size $D>6$m for detection of molecular signal in a twin Earth orbiting a GK star brighter than $V=5$, which has never been reported yet and is expected to be very rare. We in the meantime initiate a CubeSat mission named Nearby Earth Twin Hunter (NETH) (Wang et al. \textit{in prep.}) to monitor several nearby bright GK stars with known transiting planets and to search for twin Earths, which may find first nearby twin Earths and provide targets for the Tianlin mission. 

On the other hand, obtaining planet spectra directly and detecting molecular signals seems to be more feasible because starlight could have been extremely suppressed down by next-generation advanced coronapgrahs or external starshade. Moreover, one would expect that the sample size of non-transiting twin Earths will become much larger than that of transiting ones in the future with the onset of CHES, Tianlin and other direct imaging surveys, given that statistically non-transiting planets are more abundant than transiting planets.

The secondary science goal is to obtain a comprehensive understanding of various types of planets and planetary systems what they are made of, how they form and evolve, and what shapes their atmospheres, by conducting an in-depth spectroscopic survey of a sample ($>100$) of nearby rocky and gas planets with unprecedented precision and accuracy. This survey will eventually lead to a revolution in our knowledge of planetary science. The corresponding sample sizes of HabEX, LUVOIR-A and -B are 150, 648 and 576, respectively~\citep{Gaudi2020, LUVOIR2019}.

Nevertheless, Tianlin will provide key capabilities for other high-impact general astrophysical questions. For example, its high sensitivity and spatial resolution in all bands, particularly the UV bands, will help to constrain the nature of dark energy and dark energy, to understand the complexity in the formation and evolution of cosmic web and galaxies, to learn in-depth the connection of the structures, dark matter, radiation, and baryons.  

The goal to explore the habitability and detect potential biosignatures of nearby twin Earths is extremely challenging, simply because an Earth-like object is extremely faint and is deeply embedded in its host star's light. These facts put strong constraints to the baseline setup of Tianlin, particularly its aperture size, noise level, and contrast level of the star-light suppression technique, as indicated by our simulations carried out in Section~\ref{sect:model}. In the rest part of this manuscript, we will mainly discuss the implications obtained from the first scientific objective, i.e., the search of biosignature from exoplanets.

\section{Model Description and methodology}
\label{sect:model}

In this section, we describe briefly what molecules are to search for, how we generate planet and star spectra, how to simulate observations, and estimate the required integration time for each configuration. More details will be presented in Zhai et al. 2023 (\textit{submitted.}) 

\begin{table}[ht]
\caption{Input parameters of the simulation, and the recommended values suggested by our simulation.}
\label{tab:simu_input}
\begin{center}       
\begin{tabular}{llll} %%
\hline
\hline
Telescope/instrument &  Parameter space & Recommended value &   \\
\hline
 Telescope aperture $D$ (m) &  [4,12]& 6\\
 Detector dark current $d$  (e$^-$/hr/pixel) & [0.02,2] & 0.2 \\
 Contrast level $C$ & [$10^{-11},10^{-7}$] & $10^{-10}$\\
 Throughput of spectrograph $\eta_{\rm spe}$ (\%) & [40, 70] & 70\\
 Throughput of coronagraph $\eta_{\rm cor}$  (\%)& [5, 20] & 10 \\
 Lifetime (years)& [5, 10] &  10\\
 Exposure time $T$ (hours) & $100-10000$ & \\
 Wavelength $\lambda$ ($\mu$m) & fixed, $0.8-1.05$ &  \\
\noalign{\smallskip}
\hline 
\end{tabular}
\end{center}
\end{table} 

%%%%% Fig:spectra_mole  %%%%%%%%%%%%%%%%%%%%
\begin{figure}
\begin{center}
\begin{tabular}{c}
\includegraphics[width=\textwidth, trim=40 20 40 20,clip]{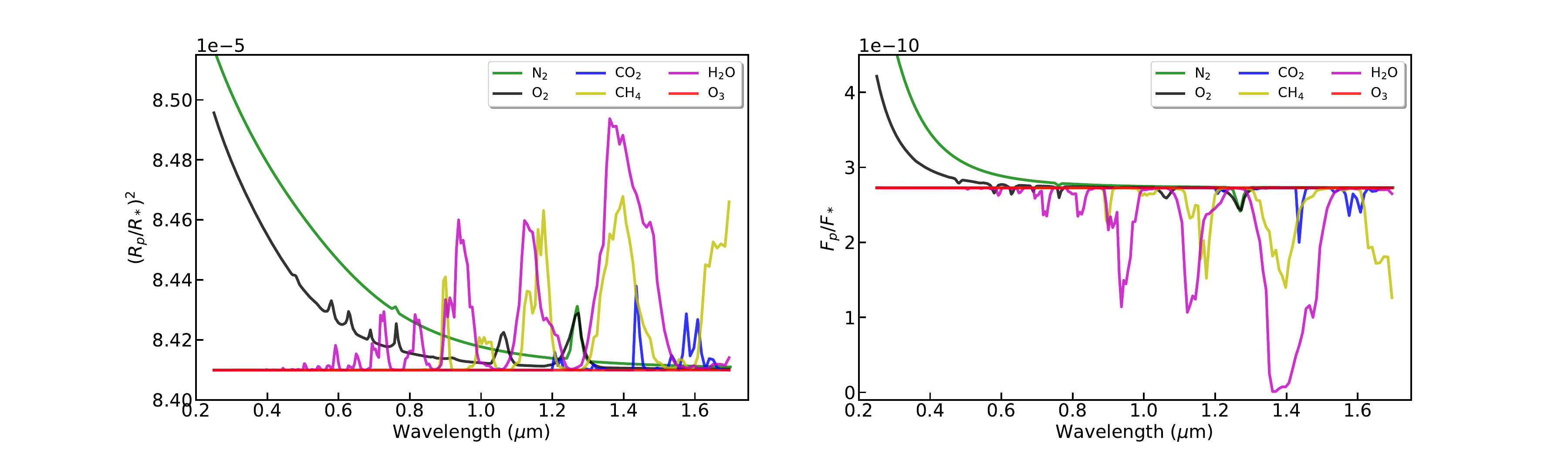}
\end{tabular}
\end{center}
\caption 
{\label{fig:spectra_mole}
The model-predicted reflection (\textit{Left}) and transmission (\textit{Right}) spectra from 0.25 to 1.7$\mu$m at $R$=150 for  N$_2$, CO$_2$, H$_2$O, O$_2$, CH$_4$ and O$_3$, respectively.} 
\end{figure} 
%%%%% end of fig:spectra_mole  %%%%%%%%%%%%%%%%%%%%

The fundamental question we attempt to address here is whether the proposed Tianlin mission concept has the capability in its lifetime to detect potential biosignature and habitability markers, and derive optimized mission requirements to achieve this goal. The two techniques that we consider employing are transmission spectroscopy (hereafter TMS) and high contrast imaging (HCI) aided by a coronagraph. 

The key signals that may provide evidence for the presence of life include gas molecules in the exoplanet atmospheres such as oxygen (O$_2$), ozone(O$_3$), methane (CH$_4$), nitrous oxide (N$_2$O), and methyl chloride (CH$_3$Cl), and substances on the surface for example the vegetation ``red edge" (VRE), halophile salt. Among them, O$_2$, O$_3$ and CH$_4$ are our primary target species to detect and study. In addition, water (H$_2$O) is one of the key life-demanding species and its non-existence may be used as an anti-biosignature. Plenty of spectral features from these four molecules are detectable in the atmospheres of rocky planets from the near UV to IR~\citep{Leger2019,Kiang2018,DesMarais2002}. We also show respectively in Fig.~\ref{fig:spectra_mole} and \ref{fig:earth_twin_spectra} the model-predicted emission and transmission spectra from individual species, and the theoretical spectrum of a twin Earth (black curve) between 0.25 and 1.7 $\mu$m at a resolution $R$ of 150, 1000 and 15000 (the up, middle and down panels), together with the simulated transmission and emission spectrum (left and right panels) for the $R=150$ case.

As the first step, the simulation performed in this work is only on the detectability of H$_2$O in the HZ Earth-like planets around GK stars using both the direct spectroscopy and transmission spectroscopy. This is because this molecule is life-demanding, and is the easiest one to detect in the optical and NIR. Detecting the crucial biosignature gas O$_2$ is harder than H$_2$O, and can be achieved simply by increasing integration time for most cases. The key technical parameters used in the simulation are presented in Table~\ref{tab:simu_input}.

\subsection{Planet spectra model}
\label{sect:planet_spectra_model}
%%%%%%%%%% Fig of T-P profile %%%%%%%%%%%%%%%%%%%%%%%%%%%%%
\begin{figure}
	\includegraphics[width=\columnwidth]{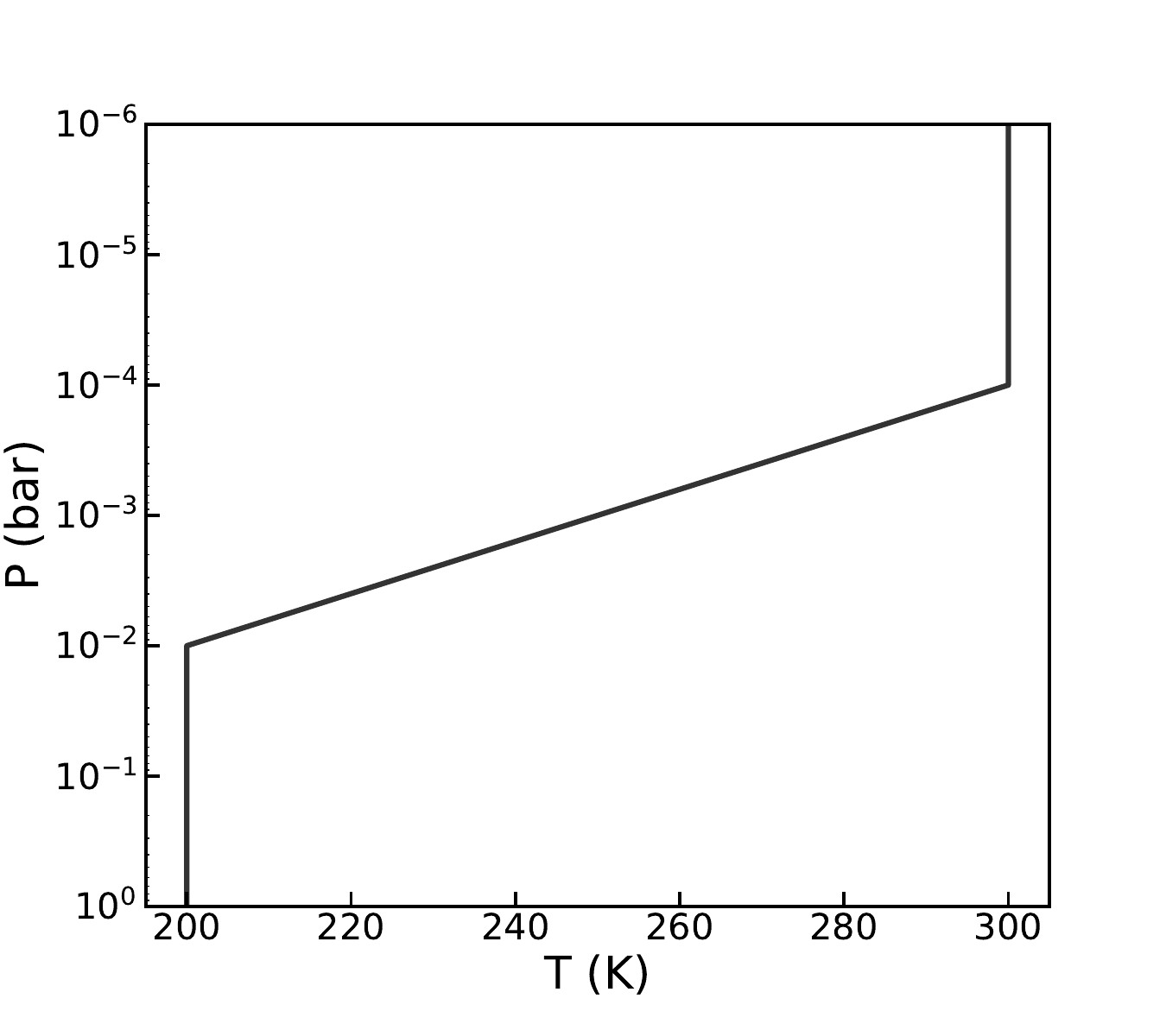}
    \caption{The T-P profile that we employ for the generation of planet spectra.}
    \label{fig:TP_profile}
\end{figure}
%%%%%%%%%% Fig of T-P profile %%%%%%%%%%%%%%%%%%%%%%%%%%%%%

In this work, the reflection and transmission spectra of Earth-like planets are generated using the public code petitRADTRANS~\citep{Molliere2019}, assuming an Earth-like surface and atmosphere. The temperature-pressure (T-P) profile we adopted is a simplified toy model with a thermal inversion layer included, as shown in Figure~\ref{fig:TP_profile}. The model atmospheres include molecules of oxygen (O$_2$), ozone(O$_3$), methane (CH$_4$), water (H$_2$O), nitrogen(N$_2$), and carbon dioxide (CO$_2$), with their volume mixing ratios (VMR) the same as those of Earth (shown in Table~\ref{tab:para_molecule}). Absorption, emission and scattering dues to H$_2$O, O$_2$, O$_3$, CH$_4$, CO$_2$, N$_2$ are taken into account. Their model-predicted emission and transmission spectral signatures are shown in Figure~\ref{fig:spectra_mole}. 
\begin{table}[ht]
\caption{VMRs of the molecules considered in the planet atmospheric  models.}
\label{tab:para_molecule}
\begin{center}       
\begin{tabular}{cc|cc|cc} %%
\hline
\hline
 Molecular &  VMR    &Molecular &  VMR  & Molecular &  VMR \\
\hline
O$_2$  & 0.21 & O$_3$  & $10^{-7}$ & CH$_4$ & 0.0001 \\
%       &      &        &      &  &        \\
%\noalign{\smallskip}&      &&      &  &        \\
\hline
%\noalign{\smallskip}
H$_2$O & $10^{-2},10^{-3},10^{-4}$ & N$_2$ & 0.78 & CO$_2$ & 0.0004\\
\noalign{\smallskip}
\hline 
\end{tabular}
\end{center}
\end{table}

The abundance of each molecule in petitRADTRANS is in unit of mass fractions, which is converted to VMR using the following formula:
\begin{equation}
X_{i}=\frac{\mu_{i}}{\mu}n_{i}
\end{equation}
where $X_{i}$ is the mass fraction of species $i$, $\mu_{i}$ is the mass of the molecule $i$, $\mu$ is 28.7, the atmospheric mean molecular weight of Earth and $n_{i}$ is the VMR of species $i$.

In total, $4\times2\times3\times2=48$ sets of high-resolution transmission and reflection spectra of rocky planets within HZs are generated, considering four types of host stars (G2, G8, K2, K7, cf. Table~\ref{tab:para_star}), two types of planets (1R$_{\oplus}$, 2R$_{\oplus}$) with the same composition bulk compositions, and three atmospheric H$_2$O mixing ratio of 0.1, 1 and 10 times of terrestrial value (i.e. dry, standard and wet, respectively), with or without water cloud. We note that the 2R$_{\oplus}$ planet may have different atmospheric composition then the Earth's, which will be amended in the future when more knowledge about such planets is obtained. The semi-major axis values tabulated in Table~\ref{tab:para_star} are the conservative inner limits of HZs corresponding to each spectral type following the work by \cite{Kopparapu2014} for G8 to K7 dwarf stars, and are exactly the Earth's semi-major axis for a G2 dwarf.

\begin{table}[ht]
\caption{Input parameters of planet host stars.}
\label{tab:para_star}
\begin{center}       
\begin{tabular}{ccccccc} %%
\hline
\hline
 Star type &  Temperature & Mass& Radius  & Semi-major axis \\
 &(K)& ($M_\odot$) & ($R_\odot$) & (AU)\\
\hline

G2  & 5778 & 1 & 1 & 1\\
G8  & 5375 & 0.78 & 0.86 & 0.71\\
K2 & 4925 & 0.69 & 0.64 &0.5\\
K7 & 4017 & 0.63 &0.6 & 0.3\\
\noalign{\smallskip}
\hline 
\end{tabular}
\end{center}
\end{table}

For the water cloud, the irregular shapes of water clouds as the condensates in the atmosphere are included, with a mean particle size of 25$\mu$m. For transmission spectra, we also calculate the spectra with a gray cloud deck at 0.1 bar for reference, where is the position of the top of troposphere.

Note that we do not include stars later than K or earlier than G, as they are not our primary targets. Actually, planets around stars with spectral type later than K have much higher signals than those around GK stars, which means that any configuration that can detect spectral signals from Earth-like planets around K star can detect those Earth-like planets around M and less massive stars assuming the planets are outside of the inner working angle (IWAs). The IWA is the smallest angular separation between host and companion sources at which the faint companion is detectable. On the other side, Earth-like planets around F-type stars have too small contrast ratios for solid detection even by a 6m space telescope with the fancy parameters.

For reflection spectra, surface albedo is set to 0.3 and phase angle to 45$^\circ$. For simplicity, we assume all the planets of consideration are beyond the inner working angles (IWAs) of our proposed coronagraph, which is $\sim$25 and 46\,mas for the UV and VIS bands respectively. The latter value corresponds to the conservative middle HZ radius of G0 and K9 stars $\sim$40 and 20\,pc away, respectively. 

\begin{figure}
\begin{center}
\begin{tabular}{c}
\includegraphics[height=14cm, trim=110 60 100 20, clip]{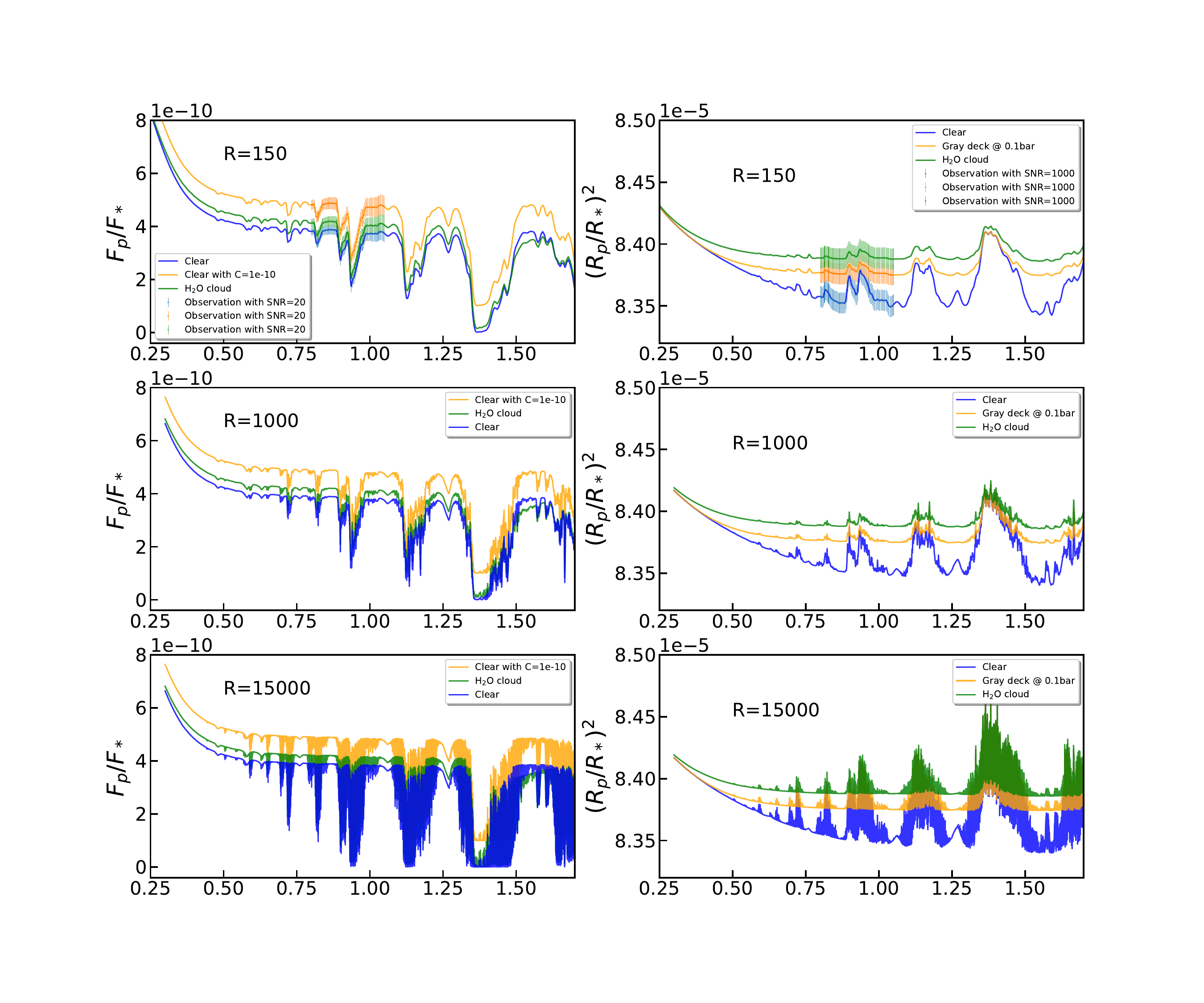}
\end{tabular}
\end{center}
\caption 
{\label{fig:earth_twin_spectra}
The simulated noise-free reflection (\textit{Left}) and transmission (\textit{Right}) spectra of an Earth twin in the NUV, optical, and NIR bands of at $R$=150 (\textit{Top}), 1000 (\textit{middle}) and 15,000 (\textit{Bottom}), respectively. Absorption, emission and scattering due to H$_2$O, O$_2$, O$_3$, CH$_4$, CO$_2$, N$_2$ are taken into account. For the emission spectra, three different cases assuming cloudy or clear and different albedos are considered. For the transmission spectra, two cases assuming cloudy or clear are included. In the top two panels, spectra with SNR(0.88$\mu$m) of 20 are overplotted for comparisons. The definition of SNR is described in Section~\ref{sect:simulated_spectra}.} 
\end{figure}

Fig.~\ref{fig:earth_twin_spectra} shows as examples the reflection and transmission spectra with or without water cloud of a twin Earth at 1\,AU orbit of a G2 star at three different R. For the reflection spectra, those in yellow include the contribution from starlight leakage with a suppression level of 10$^{-10}$ for comparisons. In each of the right panel, the yellow transmission spectrum illustrates the case of a twin Earth atmospheres with gray cloud deck presented at 0.1\,bar, i.e., the atmosphere below this deck cannot be probed due to either a large absorption opacity or atmospheric diffraction.

%%%%%%%%%%%%%%%% Fig of earth_twin_spectra %%%%%%%%%%%%%%%%%%
%%%%%%%%%%%%%%%% Fig of earth_twin_spectra %%%%%%%%%%%%%%%%%%

\subsection{Simulated spectra}
\label{sect:simulated_spectra}

An observed planet reflection spectrum $F_{\rm p}^{\rm R,Obs}(\lambda)$ is
simply the addition of planet spectrum, stellar spectrum escaped from a stellar
light suppression instrument, and noise, while a transmission planet spectrum
$F_{\rm p}^{\rm T,Obs}(\lambda)$ consists of model planet spectrum and noise,
assuming stellar contribution has been removed clearly by contrasting the
in-transit spectrum with out-of-transit spectrum. The leakage stellar signal is
estimated by simply multiplying the stellar spectra by the starlight
suppression level $C$. 

As described in Section~\ref{sect:planet_spectra_model}, the model planet
spectrum is generated by petitRADTRANS, while the synthetic stellar spectrum
$F_{\ast}^{\rm Mod}(\lambda)$ is obtained from PHOENIX
library~\citep{Husser2013}. Both spectra are then convolved with a Gaussian
profile that corresponds to a certain spectral resolution $R$. Then Poisson
noise is added to the convolved spectra assuming a Gaussian distribution,
assuming a certain SNR at the reference wavelength at 0.88$\mu$m and any other
wavelength of interest, (cf. definition of SNR and SNR($\lambda$) in
Section~\ref{sect:SNR}). In the top panels of
Fig.~\ref{fig:earth_twin_spectra}, the simulated observation spectra with
SNR=20 for reflection spectra and SNR=1000 for transmission spectra from 0.8 to
1.05 $\mu$m are overplotted with error bars as examples. 

Although Tianlin is proposed to cover a wavelength range of 0.2 to 1.7$\mu$m,
here in this work we use only the $0.8-1.0\mu$m part for the simulation. The
main purpose of this simulation is to verify whether water feature can be
detected in a specific star-planet system and what is the required integration
time. Although water has strong features in NIR, but the current available NIR
detectors that may be employed for Chinese missions may probably bear large
readout noise and dark current, making it more difficult for detecting water
features in NIR than in the optical. In the optical, the strongest water
features reside in the 0.8-1.0$\mu$m range, thus including other bands will not
enhance signals significantly. On the other hand, enlarging wavelength coverage
will require much more time for the spectral information retrievals,
particularly for high-R cases. Detailed simulation with wider spectral coverage
and on other species will be performed in the future.

In summary, the observed reflection and transmission spectra are given respectively by:
 \begin{equation} F_{\rm p}^{\rm R,Obs}(\lambda) = F_{\rm p}^{\rm R,Mod}(\lambda)+F_{\ast}^{\rm Mod}(\lambda)\cdot C+P(\lambda),
 \end{equation} 
 and
\begin{equation} F_{\rm p}^{\rm T,Obs}(\lambda) = F_{\rm p}^{\rm T,Mod}(\lambda)+P(\lambda),
\end{equation} 
where $P(\lambda)$ is the added Poisson noise, at a given SNR.

\subsection{Molecule constraint definition}
\label{sect:mol_detection}

The key of this simulation is to detect H$_2$O in the simulated observation, thus its mixing ratio is of primary interest. Following the same method of analyzing real observed planet spectra, a spectral retrieval technique is employed to infer from the observed spectra the temperature structure and chemical compositions of the target planet. We define it as a \textit{constraint} if a posterior shows a localized peak and has a 1$\sigma$ range of less than 1/5 of retrieved posterior values. For H$_2$O in our simulation, the input of the mixing ratio is $-3.2$ in logarithmic scale, and thus the 1$\sigma$ range is less than 0.6 orders of magnitude. Our criteria of constraint are thus stronger than those defined in \cite{Feng2018} and \cite{Smith2020}, in which one order of magnitude 1$\sigma$ range will define a constraint.

The mixing ratio of a certain molecule is thus retrieved using the Nested Sampling Monte Carlo package PyMultiNest~\citep{Buchner2014} of petitRADTRANS from the simulated observed spectra with a certain SNR and spectral resolution $R$. In this way, the lowest SNR values to achieve for a constraint to H$_2$O VMR are then determined for each spectrum at given $R$. 

\subsection{SNR and total integration time}
\label{sect:SNR}

We use the noise model of \cite{Robinson2016} for the estimation of SNR that could be achieved by the given instrument. The desired SNR values at a certain $\lambda$ are connected to total integration times $t{_{\rm int}}$ for a set of technical parameters and a given noise model. Following Equation 6 of \cite{Robinson2016}, assuming a background subtraction will be performed to detect the planet signal, the total integration time $t{_{\rm int}}$ required for a reflection spectral to reach a given SNR is related to background count rate, planet photon count rate, i.e.,  
\begin{equation}
\label{eq:tint_emi}
 t_{\rm int}=\frac{n_{\rm p}+2n_{\rm b}}{n_{\rm p}^2}{\rm SNR}^2,
\end{equation}
where $n_{\rm p}$ is the planet photon count rate, $n_{\rm b}=n_{\rm z}+n_{\rm ez}+n_{\rm lk}+n_{\rm D}+n_{\rm R}$ is the total background count rate, $n_{\rm z}$ is local zodiacal light count rate, $n_{\rm ez}$ is exozodiacal light count rate, $n_{lk}$ is the leakage starlight photon count rate from the coronagraph, and $n_{D}$ is dark noise rate. As our observation shall normally stare on one target for long exposures, the detector readout noise rate is significantly smaller than the other noise and thus is not taken into account in this study. The zodiacal and exozodiacal background count rates are given using Equations 8\,\&\,9 and the quoted V band surface brightness in \cite{Stark2019}. More detailed calculations of each noise can be found in \cite{Robinson2016}. For example, for the case of an Earth-twin around a $V=5$ Sun-like star, when $C=10^{-9}$, $\eta=0.1$, $D=6$\,m, we have $n_{\rm p}=2.85\times10^{-3}$\,s$^{-1}$, $n_{\rm z}=5.66\times10^{-2}$\,s$^{-1}$, $n_{\rm ez}=1.42\times10^{-1}$, $n_{\rm lk}=7.38\times10^{-3}$\,s$^{-1}$, then $n_{\rm b}=0.21$\,s$^{-1}$. If we require a SNR of 19.8, we need a $t_{\rm int}=5.64\times10^{3}$\,hrs. The corresponding SNR$_{\rm mol}$ (as described below) for water is 3.

For transmission spectra, as photon noise from host stars vastly dominate against background noises, SNR can be accurately estimated by $n_{\rm p}/n_{\ast}^{0.5}$, where $n_{\ast}=\eta F{_\ast}t_{\rm int}$ and $n_{\rm p}=\eta F{_\ast}t_{\rm int}\Delta(R_{
\rm p}/R_{\ast})^2$. Thus we have
\begin{equation}
\label{eq:tint_trans}
 t_{\rm int} = \frac{{\rm SNR}^2}{\eta F_{\ast}\Delta(R_{\rm p}/R_{\ast})^2},
\end{equation}\
where $\Delta(R_{\rm p}/R_{\ast})^2$ stands for the transmission spectral signal of the tracing molecule.
Note that currently, we use a wavelength-independent end-to-end throughput $\eta$ for our simulation. 

It is obvious that the SNR is $\lambda$-dependent. We set the reference point of SNR at $\lambda_0$=0.88 $\mu$m, just outside of the water band. Considering the integration time is constant in a single bandpass, SNR($\lambda$) can be derived using the following two equations for reflection and transmission spectra

\begin{equation}
SNR(\lambda)^2\frac{n_{p}(\lambda)+2n_{b}(\lambda)}{n_{p}(\lambda)^2}=SNR(\lambda_0)^2\frac{n_{p}(\lambda_0)+2n_{b}(\lambda_0)}{n_{p}(\lambda_0)^2},
\end{equation} 
and
\begin{equation}
SNR(\lambda)^2\frac{n_{\ast}(\lambda)}{n_{\rm p}(\lambda)^2}=SNR(\lambda_0)^2\frac{n_{\ast}(\lambda_0)}{n_{\rm p}(\lambda_0)^2}.
\end{equation}

Note that the total integration time required to achieve a given SNR is different from the SNR$_{\rm mol}$, which is the minimum SNR required to detect a certain molecule as defined for example by~\cite{BS2014}. The main difference between SNR and SNR$_{\rm mol}$ is the definition of the signal. The signal involved in SNR$_{\rm mol}$ is the molecular spectral feature imprinted on the continuum, i.e., the difference between the continuum level and the valley value (or the peak value) for the case of emission (or transmission) spectrum. For a given planet spectrum, the two SNRs are numerically connected. Taking the R=150 clear emission spectrum as an example (the blue curve in the top-left of Fig.~\ref{fig:earth_twin_spectra}), the signal  of SNR is $F_{\rm p}/F_\star$ and equals to  $\sim4\times10^{-10}$ at 0.88$\mu$m , while the signal of SNR$_{\rm mol}$ for H$_2$O is $F_{\rm p}/F_\star(0.88\mu{\rm m})-F_{\rm p}/F_\star(0.94\mu{\rm m})$   $\approx 4\times10^{-10}-2\times10^{-10} = 2\times10^{-10}$. So, for this case, the SNR at 0.88$\mu$m is about twice of SNR$_{\rm H_{2}O}$. Based on this noise-free spectrum, we generate a series of noisy spectra with different SNR$_{\rm H_{2}O}$, which are used to determine the abundance of H$_2$O via the retrieval analysis. The smallest SNR$_{\rm H_{2}O}$ that allows a ``constraint'' of H$_2$O abundance is the SNR$_{\rm mol}$, and the SNR(0.88$\mu$m) can be calculated from the corresponding noisy spectra. The last step is to calculate $T_{\rm int}$ using Eq.~\ref{eq:tint_emi}.

\section{Results and Discussion}
\label{sect:result}

\subsection{Direct spectroscopy}
Our simulation suggests that under the optimized instrumental performance, Tianlin is expected to yield emergent spectra with SNR$_{\rm mol}=5$ for H$_2$O with its coronagraph after integrating $10^2-10^3$ hours for the vast majority of the standard and wet Earth-like planets in the HZs of G8-K7 stars, assuming the phase angle of 90$^\circ$. Increasing $D$ to 6m will make Tianlin capable of measuring the water vapor of a nearby twin Earth in less than $10^3$~hrs. 
 
\begin{figure}
\begin{center}
\begin{tabular}{c}
\includegraphics[width=\columnwidth]{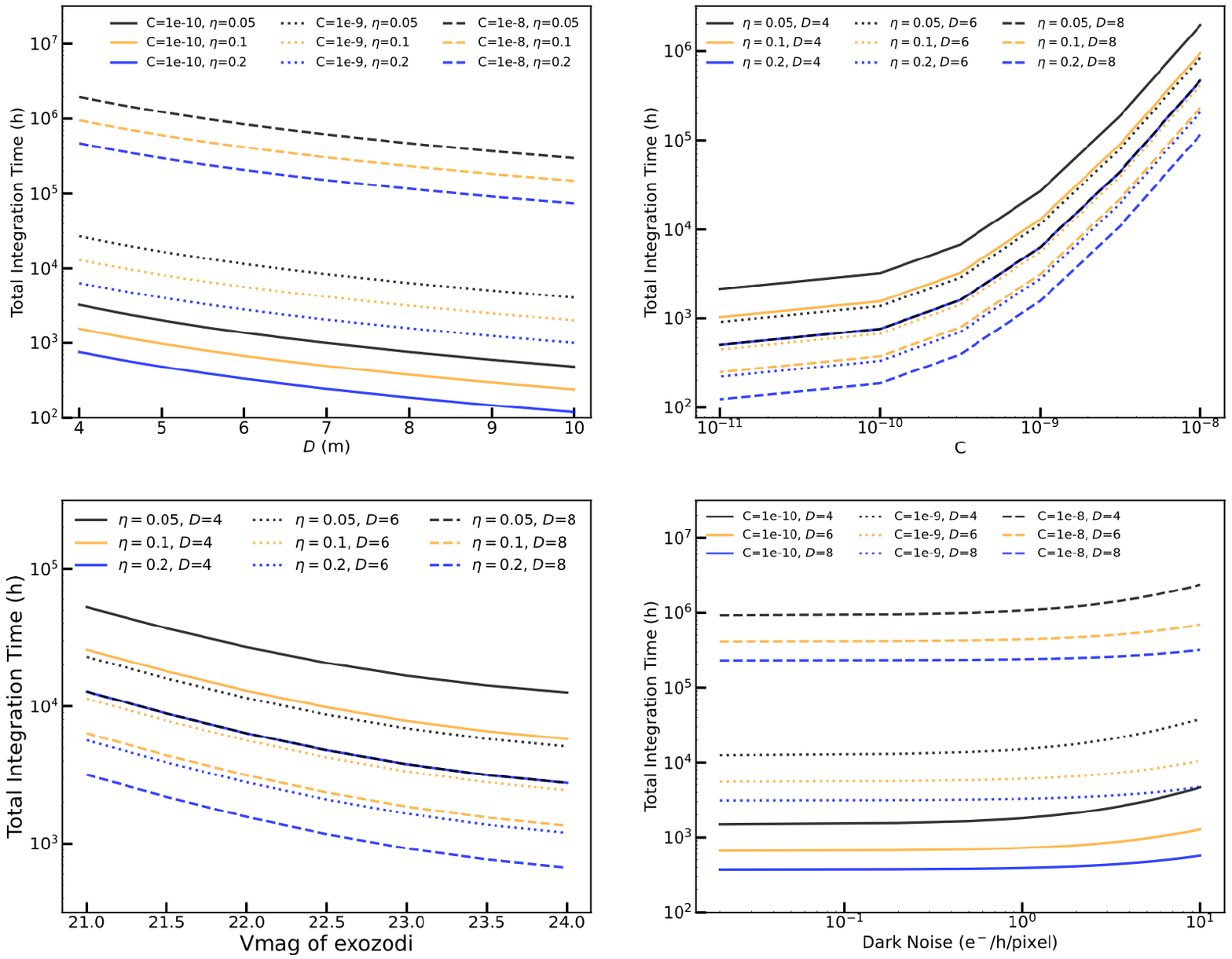}
\end{tabular}
\end{center}
\caption 
{\label{fig:instrument}
{\textit Left:} The dependence of the total integration time $t_{\rm int}$ on
the aperture size $D$ for the direct spectroscopy technique. $t_{\rm int}$
is the minimum integration time for a constraint of water vapor on a twin
Earth around a $V=5$ G2V star, based on our simulation of $R=150$ reflection
planet spectra. Lines with different styles and colors represent difference
$t_{\rm int}-D$ dependence with different $C$ and $\eta$. \textit{Right:} The
dependence of $t_{\rm int}$ on $C$ with different combinations of $\eta$ and
$D$.}
\end{figure}

The main results of our simulation on reflection spectra are presented in Fig.~\ref{fig:instrument}, which shows the dependence of $t_{\rm int}$ on $D$, $C$ and $\eta$ for a twin Earth with the Earth-like water vapor content.
The dependence on dark current rate $d$ is very weak until $d$ exceeds 2e$^-$/h/pixel, and even lower dark current is already accomplished by modern detector developing companies such as Teledyne e2v\footnote{https://www.teledyne-e2v.com/}. This is reasonable because in this regime, noise is dominated by local zodiacal and exozodiacal light backgrounds.

It is clear from Fig.~\ref{fig:instrument} that $t_{\rm int}$ is very sensitive to $C$, the contrast of a coronagraph or an external starshade, especially when $C>10^{-9.5}$. It must be better than $10^{-9}$ to constrain H$_2$O of a twin Earth in a maximum affordable integration time, i.e., less than 10$^4$ hours, or about 1 year. Therefore, the starlight suppression performance, particularly the contrast $C$ is critical for the achievement of Tianlin's objective. Such limits can be achieved in a laboratory experiment with a raw contrast $10^{-9}$ or better at visible wavelengths~\citep{Trauger2007,Dou2016,Prada2019,Llop-Sayson2020,Harness2021}, and the improved post-processing analysis could further reduce residual starlight speckles. It will be certainly more challenging to realize such a high contrast on board a spacecraft, mainly due to fast line-of-sight jitter and slow wavefront drift. As shown in \citet{Gaudi2020}, the self-induced pointing jitter for HabEx-like telescope at the L2 point with tiny thermal and gravitational gradient disturbances should be nonexistent, while the reaction wheels or microthrusters may induce pointing jitter in the order of 10$^{-4} {''}$ rms. The simulation done for HabEx shows that such jitter won't affect the raw contrast at IWA to the level of $3\time10^{-10}$. On-board testing in space for $>10^8$ starlight suppression will be carried out on the near-future Nancy Grace Roman Space Telescope~\citep{Akeson2019WFIRST}, and $\sim10^8$ contrast ratio on the Chinese Space Station Telescope (cf. CSST whitebook, \textit{private communication}). 

In addition, Fig.~\ref{fig:instrument} suggests that the second key technical parameter $D$ should not be smaller than 6m. It requires more than 1550 and 670 hours for $D=4$m or 6m, assuming $C=10^{-10}$ and $\eta=0.1$. These two numbers increase dramatically to 13,000 and 5600 hours for $C=10^{-9}$. Therefore, a primary mirror with $D=4$m seems not large enough to secure robust detection of atmospheric biosignatures from Earth-like planets, especially in the cases that the planet is smaller than the Earth, or the star is larger and brighter than the Sun, or the planet system is more than 10\,pc from us. Enlarging $D$ to 6~m will result in a gain in photons by a factor of $\sim$2, which will consequently increase largely the number of planets that can be observed and characterized. It is also shown that the higher $\eta$ is, the shorter $t_{\rm int}$ is required. Increasing $\eta$ seems to have a similar effect as increasing $D$, because both parameters are directly related to the number of photons received.

\subsection{Transmission spectroscopy}
\begin{figure}
\begin{center}
\begin{tabular}{c}
\includegraphics[height=9cm, trim=10 10 10 10, clip]{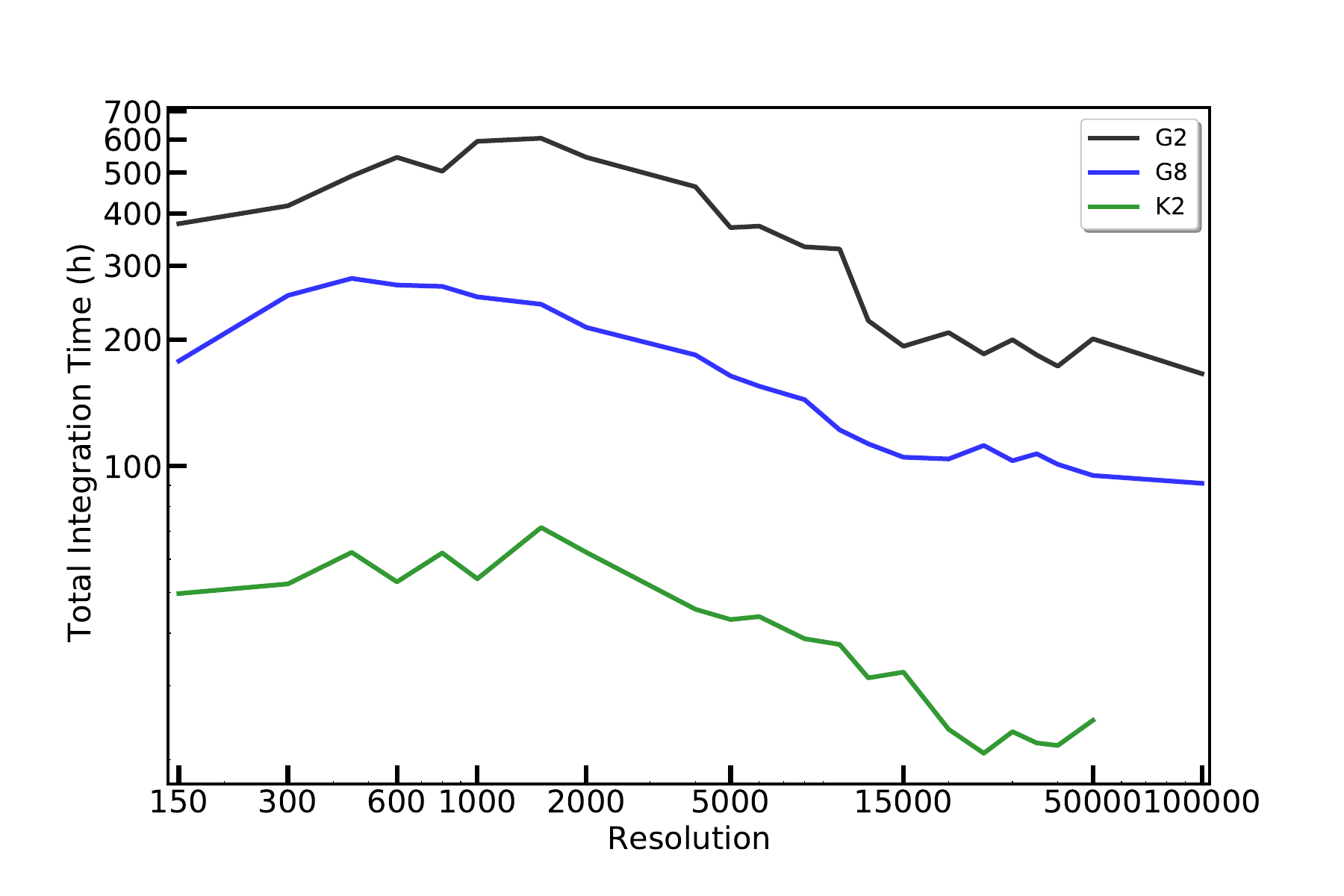}
\end{tabular}
\end{center}
\caption 
{\label{fig:TintvsR}
The required total integration time for a 6m space telescope to obtain a
constraint of water feature at 0.94$\mu$m varying with spectral resolution $R$
for an Earth-like planet orbiting a dwarf star of $V=5$ with a spectral type of
G2, G8 or K2, respectively, assuming the noise follows Gaussian distribution.
Note that this is the total integration time during primary transits, which
is $\sim$13 hours per year for a Sun-Earth system, and is $\sim$65 hours for a
5 year mission lifetime.}
\end{figure}

On the other hand, our results show that using classic low resolution ($R\sim 150$) transit spectroscopy, much longer $T_{\rm int}$ is required to achieve the same SNR$_{\rm mol}$. For example, as shown by the starting points of the lines in Fig.~\ref{fig:TintvsR}, a 6m telescope needs to integrate $\sim50$ and $>180$ hours to constrain H$_2$O VMR at 0.94$\mu$m for an Earth-sized planet orbiting a K7 and K2-G2 dwarf star $\sim$10\,pc away, respectively. Unfortunately, for transit spectroscopy, the total usable integration time is strictly limited by the number of transits that could be covered in the mission's lifetime and the duration of each transit, with the latter about 13 hrs for the Sun-Earth system. For Tianlin, the conservative mission lifetime is 5 years, with possible 5 years extension. Therefore, the total integration time is approximately 65 and 130\,hr, for a habitable planet around a G2 and K7 star, respectively. Increasing $D$ to 8m will significantly shorten the required $T_{\rm int}$, and thus make it possible to detect H$_2$O with SNR$\sim$5 on an Earth-like planet in the HZ of K type star in mission lifetime, however for the same case around G types stars, one may need $D\geq$8m to detect H$_2$O. 

%%%%% The figure showing the relationship between Tint and R %%%%%%%%
%%%%% end of fig:instrument  %%%%%%%%%%%%%%%%%%%%

Lastly, we have studied the advantages of using high-$R$ technique for transit spectroscopy. Our simulations show that increasing $R$ will significantly shorten the required $T_{\rm int}$ time, at least for nearby bright stars. As shown in Fig.~\ref{fig:TintvsR}, for a constraint of water content on an Earth-like planet around a G2 star at 10\,pc, Tianlin with $D$ of 6m demands $T_{\rm int}\sim$380, 600 and 210 hours with R=150, 1500 and 15,000, respectively. Increasing R will lead to two contrary changes of SNR and thus $t_{\rm int}$. On one hand, the molecular signal per pixel decreases linearly with increasing $R$, while the detector noise does not change, and thus the SNR will decrease proportionally to the square root of $R$ approximately. On the other hand, when $R$ increases, the total integrated signal of many individual lines of a molecule will first increase quickly because the mixed individual lines at low $R$ will gradually separate from each other, and the peak intensity of each line will increase as well. Nevertheless, the increase will turn slow and eventually stop after all individual lines are separated and the instrumental lines' width becomes smaller than the intrinsic widths of planet lines. The combination of these two effects results in a decrease and then an increase of $t_{\rm int}$ when $R$ changes from $\sim$100 to $\sim$1000, followed by a steady decrease to $R\sim$15000 for G2 and G8 host star, and to $R\sim$20000 for K2 host stars. This trend is reasonable given that the median separation of every two adjacent water lines in the wavelength range from 0.8 to 1.05$\mu$m is found to be 0.6\,AA,  corresponding to a resolution of $\sim13000-16000$. Specifically, when $R=150$, only 0.07\% lines could be separated well from their neighbours and thus resolved. This fraction only increases to 1.9\% for $R=1000$, and is 53\% and 61\% for $R=15000$ and 20000, respectively.

%%%%% The figure showing the relationship between Tint and R %%%%%%%%
%%%%% The figure showing the relationship between Tint and R %%%%%%%%

We point out one important bonus advantage for the case of $R>15000$ which has not been taken into account in the above analysis. The molecular signals from individual lines can be combined by cross-correlating the observed spectra with a model template spectrum. In this way, the significance of a molecular species is subsequently enhanced by the square root of $N_{\rm lines}$, which is a multiplication factor that takes into account the number and strength of the individual planet lines involved~\citep{Snellen2015}. For the 0.94$\mu$m H$_2$O band, we find that $N_{\rm lines}\geq100$. Therefore, the required $T_{\rm int}$ in theory for a twin-Earth should be less than 100 hours using the cross-correlating technique. Simulation will be performed in the future to confirm this number.

The high-resolution spectroscopy can be also used in conjunction with the high contrast imaging technique as well to achieve a contrast ratio of $10^{-10}$ in ideal cases~\citep{Snellen2015,WangJ2018}. More realistic simulation will be further performed in order to identify the $R$ value which is the best for detecting molecular signals, and in the meanwhile can be realized in a space telescope at reasonable expenses.
\\
\section{Summary}
\label{sect:summary}
To summarize, the Tianlin mission is a space telescope with a 6+ meter aperture initiated in China with the primary objective to detect biosignatures of Earth-like planets in the HZ of nearby GK dwarf stars. Its design will be optimized for high-contrast imaging aided by an advanced coronagraph and medium-to-high resolution spectroscopy with a light, compact and stable spectrograph. In addition, Tianlin may yield a comprehensive understanding of the characteristics, formation, and evolution of planets by conducting a complete survey of gas giant planets to rocky planets orbiting nearby stars. Moreover, Tianlin will provide tremendously higher quality data for the studies of almost all astrophysical fields, including cosmology, galaxies, stars and interstellar medium, and should largely lift up our understanding of the universe and its contents.

A preliminary concept study has been conducted in order to yield the optimal and tolerable values for a set of key technical parameters for the telescope and its instrumentation. The study is based on a set of simulated planet spectra generated using petitRADTRANS, taking into consideration various stellar and planet parameters and a range of technical parameters. We conclude that to achieve our primary science goal, $D$ is recommended to be no less than 6m,  $C$ should be better than 10$^{-9}$, and $\eta_{\rm cor}$ larger than 0.1, assuming a classic internal coronagraph rather than an external starshade will be deployed. We find that the current $\lambda$ range is good enough for our purpose, and will explore other $\lambda$ ranges for smaller $t_{\rm int}$. We point out that increasing spectral resolution $\sim15,000-20,000$ may largely improve the detection capability of molecular signals with transmission spectra, similar to the case of reflection spectra, in the sense that combing high spectral resolution with high contrast imaging could significantly enhance the capability of detecting tiny signals. Extending wavelength coverage to 0.20$\mu$m or even shorter, will not only allow the detection of biosignature O$_3$, but also  enable great developments in various studies including the cosmic web, high-energy physics in stars and galaxies, for which EUV is the best window however ignored for now and in the coming decades. Additional in-depth and more realistic simulations are ongoing to further constrain and better refine the baseline configuration of Tianlin, which will be followed by a feasibility study in the coming years.
\normalem

\begin{acknowledgements}
This research is supported by the National Natural Science Foundation of China (NSFC) grants Nos.~11988101, 42075123, 42005098 and 62127901, the National Key RD Program of China No.~2019YFA0405102, the Strategic Priority Research Program of Chinese Academy of Sciences (CAS), Grant No~.XDA15016200 and XDA15072113. WW is supported by the Chinese Academy of Sciences (CAS), through a grant to the CAS South America Center for Astronomy (CASSACA) in Santiago, Chile. We acknowledge the science research grants from the China Manned Space Project with NO. CMS-CSST-2021-B12.

\end{acknowledgements}
  
\bibliographystyle{raa}
\bibliography{bibtex}

\end{document}